\documentclass[%
 reprint,
 superscriptaddress,
nofootinbib,
 amsmath,amssymb,
 aps,
]{revtex4-2}

\usepackage{graphicx}
\usepackage{dcolumn}
\usepackage{bm}
\usepackage{hyperref}

\usepackage{physics}

\begin{document}

\preprint{APS/123-QED}

\title{Deciphering the Scattering of Mechanically Driven Polymers using Deep Learning}
\author{Lijie Ding}
\affiliation{Neutron Scattering Division, Oak Ridge National Laboratory, Oak Ridge, TN 37831, USA}
\author{Chi-Huan Tung}
\affiliation{Neutron Scattering Division, Oak Ridge National Laboratory, Oak Ridge, TN 37831, USA}
\author{Bobby G. Sumpter}
\affiliation{Center for Nanophase Materials Sciences, Oak Ridge National Laboratory, Oak Ridge, TN 37831, USA}
\author{Wei-Ren Chen}
\affiliation{Neutron Scattering Division, Oak Ridge National Laboratory, Oak Ridge, TN 37831, USA}
\author{Changwoo Do}
\email{doc1@ornl.gov}
\affiliation{Neutron Scattering Division, Oak Ridge National Laboratory, Oak Ridge, TN 37831, USA}

\date{\today}

\begin{abstract}

We present a deep learning approach for analyzing two-dimensional scattering data of semiflexible polymers under external forces. In our framework, scattering functions are compressed into a three-dimensional latent space using a Variational Autoencoder (VAE), and two converter networks establish a bidirectional mapping between the polymer parameters (bending modulus, stretching force, and steady shear) and the scattering functions. The training data are generated using off-lattice Monte Carlo simulations to avoid the orientational bias inherent in lattice models, ensuring robust sampling of polymer conformations. The feasibility of this bidirectional mapping is demonstrated by the organized distribution of polymer parameters in the latent space. By integrating the converter networks with the VAE, we obtain a generator that produces scattering functions from given polymer parameters and an inferrer that directly extracts polymer parameters from scattering data. While the generator can be utilized in a traditional least-squares fitting procedure, the inferrer produces comparable results in a single pass and operates three orders of magnitude faster. This approach offers a scalable, automated tool for polymer scattering analysis and provides a promising foundation for extending the method to other scattering models, experimental validation, and the study of time-dependent scattering data.

\end{abstract}

\maketitle


\section{Introduction}

Polymers are fundamental components in both nature and industry,\cite{de1979scaling, de1990introduction} serving as the backbone for a wide range of materials—from biological tissues\cite{mohanty2000biofibres,mohanty2002sustainable} to high-performance synthetic composites.\cite{oladele2020polymer} Their omnipresence is matched by the diverse roles they play, whether in maintaining structural integrity,\cite{wang2020self, priftis2009surface, xia2021review} facilitating energy transfer,\cite{hager2020polymer, wang2022progress} or enabling advanced technological applications.\cite{ilami2021materials, herzberger2019polymer} In many real-world settings, materials encounter external forces or flow conditions,\cite{ghanem2021role,wang2011salient,smith1999single} underscoring the importance of understanding how polymers respond under such circumstances. The mechanical response of polymers not only reveals their intrinsic molecular dynamics but also governs critical properties such as strength, flexibility, and durability. Investigating these responses allows us to gain valuable insights that inform the design and optimization of new materials. 

Scattering techniques, especially small-angle scattering (SAS) such as neutron and X-ray scattering are indispensable for probing the internal structure of polymers\cite{chu2001small} at the nanoscale. These methods offer detailed insights into polymer conformation, phase behavior, and the distribution of molecular segments, thereby bridging the gap between macroscopic mechanical properties and microscopic structural dynamics. When combined with rheometer, RheoSAS\cite{murphy2020capillary,eberle2012flow,bharati2019poiseuille} allows us to study the behavior of polymer under flow conditions, capturing how the external forces modify the conformation of the polymer chain. Such information is critical for linking the molecular structure of polymers to their performance in real-world applications, serve as powerful tools for guiding the design and optimization of new polymer materials. However, despite these strengths, the current implementation of RheoSAS faces notable limitations. In particular, it often falls short in quantitatively correlating the effective force experienced by the polymer chains with the macroscopic flow conditions.\cite{eberle2012flow} This disconnect makes it difficult to fully predict how microscopic chain dynamics influence the overall mechanical performance of the material under practical processing and service conditions.

Recent developments in machine learning (ML)\cite{murphy2012machine,carleo2019machine} and deep learning\cite{goodfellow2016deep,lecun2015deep} offer some promising methods to map the scattering function of the polymers to the effective forces acting on them under flow conditions. Such mapping relies on data generated by high-quality physical simulation of the polymer chains, which are then used to train the machine learning models. Two primary approaches have been developed, one employs Gaussian Process Regression (GPR)\cite{williams2006gaussian, chang2022machine} to directly correlate the scattering function with system parameters, which has been applied to colloids\cite{ding2024shearcolloid} and polymer\cite{ding2024machine,ding2024ladder,ding2025charge} systems. Another leverages neural networks to construct a generative model that can produces the scattering function on the fly based on the system parameters. This generative model can then be integrated with a least-squares fitting algorithm to extract parameters from measured scattering function, a technique previously applied to one-dimensional scattering functions of colloids\cite{tung2024inferring,tung2024scattering} and lamellar\cite{tung2024unveiling,tung2025insights} systems. Although the GPR approach is relatively straightforward and efficient in parameter extraction, its training often requires manual tuning in hyperparameters finding, and become slow for large data set. In contrast, while the neural network approach is more generalized and automated, it is hindered by the extensive number of iterations needed during the least-squares fitting process. 

In this work, we develop a Variational Autoencoder\cite{kingma2013auto} (VAE)-based neural network approach that allow us to construct both a generator for scattering function prediction and an inference network that directly extracts polymer parameters from scattering data. We then apply this combined framework to analyze two-dimensional scattering functions of polymers under external forces. We begin by generating a data set using an off-lattice Monte Carlo (MC) simulation for mechanically driven polymers \cite{ding2024off}, where the polymer energy depends on three parameters: bending modulus, stretching force, and shear. Next, we train our neural network—including the VAE module—on this simulation data. Finally, we test the performance of our deep learning model on separate test data, demonstrating the versatility and practicality of our approach.

\section{Method}

\subsection{Monte Carlo Simulation}
We use the previously developed off-lattice MC simulation\cite{ding2024off} to sample the configuration space of the polymer, which is modeled as a chain of N connected bonds with fixed length $l_b$, This continuous model produces accurate mechanical responses and avoid the orientational bias inherent in lattice models.\cite{tung2024discretized} Two types of non-local moves—namely, the crankshaft and pivot moves—are employed to update the polymer configuration. The crankshaft move randomly rotates an internal sub-chain of the polymer, whereas the pivot move rotates a sub-chain that includes one of the polymer’s ends. The tangent of bond $i$ is defined as $\mathbf{t}_i \equiv (\mathbf{r}_{i+1} - \mathbf{r}_i) / l_b$, where $\mathbf{r}_i$ denotes the position of the joint connecting bonds $i-1$ and $i$. One end of the polymer is fixed at the origin. The polymer energy is given by 
\begin{equation}
    E = \sum_{i=0}^{N-2} \frac{\kappa}{2}\frac{(\mathbf{t}_{i+1} - \mathbf{t}_i)^2}{l_b} - \sum_{i=0}^{N-1} (\gamma z_i + f)(l_b \mathbf{t}_i \cdot \mathbf{x})
    \label{equ:energy}
\end{equation}
where $\kappa$ is the bending modulus, $f$ is the stretching force applied in the $x$-direction, $\gamma$ is the shear ratio along the $z$-direction, $z_i = \mathbf{r}_i \cdot \mathbf{z}$ is the $z$-component of the position of joint $i$, and $(\mathbf{t}_i \cdot \mathbf{x})$ is the $x$-component of the bond tangent $\mathbf{t}_i$. Additionally, a hard sphere interaction with a sphere radius of $l_b/2$ is implemented between polymer joints to account for the self-avoidance. We calculate the ensemble average of the intra-polymer structure factor in the $xz$ plane $I_{xz}(\vb{Q}) = I(Q_x,Q_y=0,Q_z)$ from the positions of all of the joints, where the $I(\vb{Q})$ which is given by\cite{chen1986small}
\begin{equation}
    I(\vb{Q}) = \frac{1}{N^2} \sum_{i=0}^{N-1}\sum_{j=0}^{N-1} e^{-i \vb{Q} \cdot (\vb{r}_i - \vb{r}_j)}
    \label{equ:2d_scattering_function}
\end{equation}
To investigate the relation between the polymer parameters $(\kappa,f,\gamma)$, we randomly generate 5,000 combination of these parameters with $\kappa\in[2,20]$, $f\in [0,0.5]$, and $\gamma L\in [0,2]$, and calculate the scattering function $I_{xy}(\vb{Q})$. With both, we construct a data set $\{I_{xy}(\vb{Q})\}$ along with the polymer parameters $\{(\kappa,f,\gamma)\}$. We also use natural unit $l_b=k_B T=1$ when representing our results.

\begin{figure*}[!hbt]
    \centering
    \includegraphics[width=0.8\linewidth]{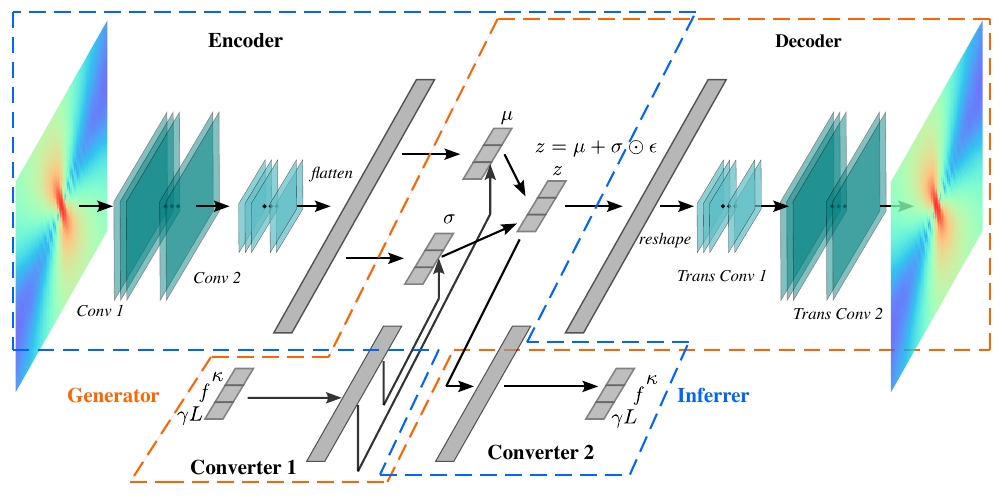}
    \caption{Architecture of the neural network that consists of a variational autoencoder (VAE) and two converters. The VAE comprises an encoder that transform the input scattering function $I_{xq}(\vb{Q})$ to the latent variable $\mu$ and $\sigma$, and a decoder that reconstructs the scattering function. Converter 1 maps the polymer parameters (bending modulus $\kappa$, stretching force $f$ and stead shear $\gamma L$) to the latent space, while Converter 2 maps the latent variables to the polymer parameters. Combining the Converter 1 with the decorder yields a generator that directly produces the scattering function from given polymer parameters, and combining the encoder with Converter 2 yields an inferrer that extracts polymer parameters from the scattering function.}
    \label{fig:NN_architecture}
\end{figure*}

\subsection{Variational Autoencoder}
To decode the two-dimensional scattering function of the polymer, we employ a VAE\cite{doersch2016tutorial} to learn its latent representation, enabling both reconstruction and parameter inference. The VAE consists of three main components: an encoder, a latent space, and a decoder. The encoder compresses the scattering function $I_{xy}(\vb{Q})$ into a three-dimensional latent representation through two convolutional layers\cite{lecun1995convolutional} followed by a linear layer, while the decoder reconstructs $I_{xy}(\vb{Q})$ from this compressed space. The two convolutional layers are of (2 stride, 32 channel) and (2 stride, 64 channel), respectively. To establish a connection between the polymer parameters  $(\kappa,f,\gamma)$ and the latent space, we introduce two additional converters. These converters, implemented as linear layers of dimension 9, facilitate bidirectional mapping between the latent variables \( \vb{z} \) and the polymer parameters  $(\kappa,f,\gamma)$ . Fig.~\ref{fig:NN_architecture} illustrates the architecture, highlighting the four key components: the VAE (encoder, decoder) and the two converters. By integrating the VAE with these converters, we achieve direct mappings between the scattering function and the polymer parameters. As shown in Fig.~\ref{fig:NN_architecture}, Converter 1, when combined with the decoder, acts as a generator, allowing the direct synthesis of scattering functions $I_{xy}(\vb{Q})$ from given polymer parameters $(\kappa,f,\gamma)$. Conversely, Converter 2, when combined with the encoder, serves as an inferrer, enabling direct extraction of  $(\kappa,f,\gamma)$  from $I_{xy}(\vb{Q})$.

The neural network is trained in three stages. In the first step, we only train the VAE using the loss function:
\begin{equation}
    L_{VAE} =\frac{1}{N} \sum_{I_{xy}(\vb{Q})} \left<\left[\log_{10}I_{xy}(\vb{Q}) - \log_{10}I'_{xy}(\vb{Q}) \right]^2 \right>_{\vb{Q}}
    \label{equ:loss_VAE}
\end{equation}
where $I'_{xy}$ is the out put of the VAE network and $\left<\dots\right>_{\vb{Q}}$ is the average over all $\vb{Q}$, and $N=|\{I_{xy}(\vb{Q})\}|$ is the number of scattering function in the data set. Then, we train the two converters. For training converter 1, we combine it with the trained decoder to form a generator and use the same VAE loss function $L_{VAE}$ to train the Converter 1 while freezing the decoder. To train the Converter 2, we combine it with the encoder to construct a inferrer and use the loss function
\begin{equation}
    L_{CVT2} =  \sum_{I_{xy}(\vb{Q})}\frac{(\kappa-\kappa')^2+(f-f')^2 + L^2(\gamma -\gamma' )^2 }{N}
    \label{equ:loss_CVT2}
\end{equation}
where $(\kappa', f',\gamma')$ are the output of the inferrer network, and the encoder is frozen during the training. Finally, we fine tune the entire network end-to-end, allow all parameters of the network to vary, using the sum of all three loss function as the total loss function.

\section{Results}
We firstly study the effect of the three polymer parameters on the conformation and scattering function of the polymer. Then, we train the neural network to encode the scattering function into the latent variables and connect the polymer parameters to the latent variables. Finally, we use the generator and the inferrer to extract the polymer parameters directly from the scattering function, demonstrate the possibility for application on experimental scattering data.

\subsection{Scattering function of the polymer}
\begin{figure}[!h]
    \centering
    \includegraphics[width=\linewidth]{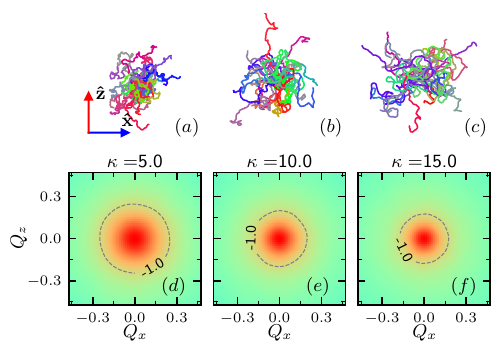}
    \caption{Configuration and scattering function of the polymer in the quiescent state for various bending modulus with $L=200$, $f=0$ and $\gamma=0$. (a)-(c) Sample configurations of the polymer chain with bending modulus $\kappa=5, 10, 15$, respectively, color indicate the end-to-end orientation. (d)-(f) Corresponding two dimensional scattering function $I_{xz}(\vb{Q})$, contour and levels represent $\log_{10}{I_{xz}(\vb{Q})}$}
    \label{fig:sample_config_Iq_kappa}
\end{figure}

\begin{figure}[!h]
    \centering
    \includegraphics[width=\linewidth]{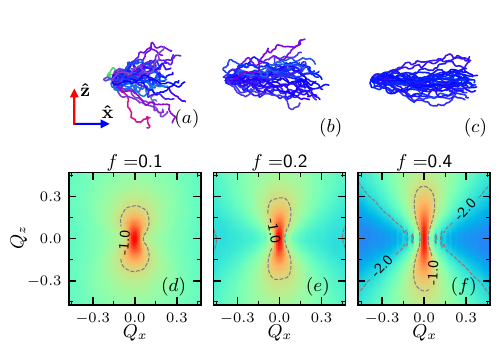}
    \caption{Configuration and scattering function of the polymer under stretch with $L=200$, $\kappa=10$ and $\gamma=0$. (a)-(c) Sample configurations of the polymer chain with stretching force $f=0.1, 0.2, 0.4$, respectively, color indicate the end-to-end orientation. (d)-(f) Corresponding two dimensional scattering function $I_{xz}(\vb{Q})$, contour and levels represent $\log_{10}{I_{xz}(\vb{Q})}$}
    \label{fig:sample_config_Iq_f}
\end{figure}

\begin{figure}[!h]
    \centering
    \includegraphics[width=\linewidth]{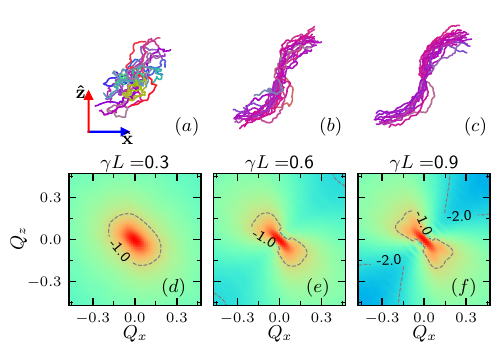}
    \caption{Configuration and scattering function of the polymer under shear with $L=200$, $\kappa=10$ and $f=0$. (a)-(c) Sample configurations of the polymer chain with stretching force $\gamma L=0.3, 0.6, 0.9$, respectively, color indicate the end-to-end orientation. (d)-(f) Corresponding two dimensional scattering function $I_{xz}(\vb{Q})$, contour and levels represent $\log_{10}{I_{xz}(\vb{Q})}$}
    \label{fig:sample_config_Iq_gL}
\end{figure}

In order to be able to extract the bending modulus $\kappa$, stretching force $f$ and steady shear $\gamma$ separately, we need to make sure they all can affect the polymer conformation and scattering function independently. Fig.~\ref{fig:sample_config_Iq_kappa}, ~\ref{fig:sample_config_Iq_f} and \ref{fig:sample_config_Iq_gL} shows the affect of the bending, stretch and shear on the conformation and scattering function, respectively. As shown in Fig.~\ref{fig:sample_config_Iq_kappa}, when the polymer is in the quiescent state with no external force, the distribution of polymer configurations is isotropic, and so is the two dimensional scattering function. Increasing the bending modulus $\kappa$ makes the polymer stiffer, which lead to more extended chain conformation. Correspondingly, the scattering function $I_{xy}(\vb{Q})$ increases at low $Q = |\vb{Q}|$, indicating stronger long-range correlation. The $I_{xy}(\vb{Q})$ are measured for $Q_x$ and $Q_y\in[-30\pi/L,30\pi/L]$ for $64\times 64$ uniformly distributed $\vb{Q}$.

When a stretching force $f$ is applied in the $\vu{x}$ direction, the polymer elongates along the that direction, while its spread in the perpendicular directions (including the $\vb{z}$ direction) decreases. Fig.~\ref{fig:sample_config_Iq_f} (a)-(c) shows variation of sample polymer configurations under the increasing stretch $f$. As the polymer extends more along $\vu{x}$ and contracts along $\vu{z}$ direction, the scattering function $I_{xy}(\vb{Q})$ decrease at low $Q_x$ and increase at low $Q_z$, forming a dumbbell shape oriented along the $\vu{z}$ direction.

Finally, as shown in Fig.~\ref{fig:sample_config_Iq_gL} (a)-(c), the steady shear $\gamma$ drive the polymer elongate towards the $\pm(\vu{x}+\vu{z})$ direction, forming part of the S-shaped configuration. Similar to the case of stretching, the scattering function $I_{xz}(\vb{Q})$ evolves into an oval or dumbbell shape, with the dumbbell oriented towards the $\pm(\vu{x}-\vu{z})$, perpendicular to the elongation direction.

\subsection{Encoding the scattering function}
\begin{figure}[!h]
    \centering
    \includegraphics[width=\linewidth]{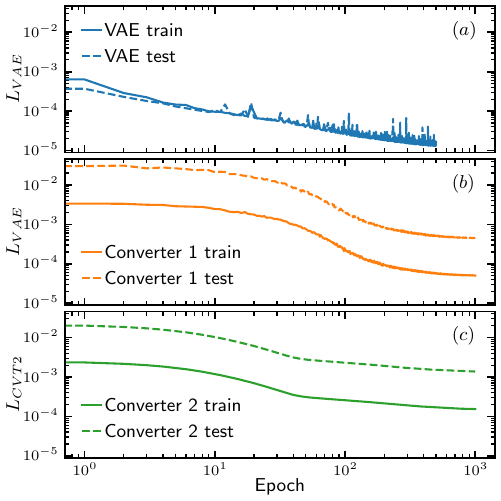}
    \caption{Training loss for the training set and test set for different training stages of the neural network. (a) Loss for the VAE when training encoder and decoder. (b) Loss for the Converter 1, which is trained while freezing the VAE. (c) Loss for the Converter 2.}
    \label{fig:loss}
\end{figure}

We first train the VAE to encode the scattering function into a three-dimensional latent space. The network is trained using the loss function $L_{VAE}$ for 500 epochs with learning rate $10^{-3}$ for the first 300 epoch and $10^{-4}$ for the rest. The Converter 1 and Converter 2 are trained for 1,000 epochs with learning rate $10^{-3}$ for the first 500 epochs and $10^{-4}$ for the rest using loss function $L_{VAE}$ and $L_{CVT2}$, respectively, ensuring accurate mapping from the latent variables to the polymer parameters. Fig.~\ref{fig:loss} shows the decay of loss function over training iteration for each training stage. The losses are evaluated by splitting the data set into a training set ($90\%$ of the data) and a test set ($10\%$ of the data). The neural network is trained exclusively on the training set, which is divided into batches of 50 samples for each parameter update.

\begin{figure}[!h]
    \centering
    \includegraphics[width=\linewidth]{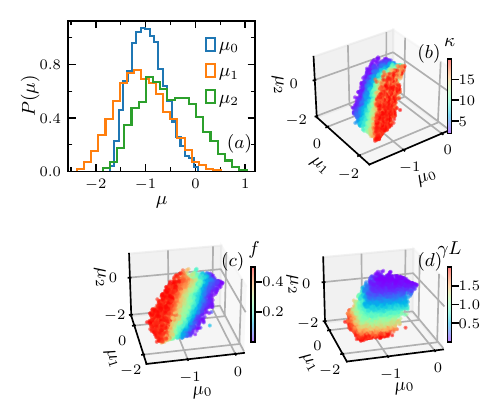}
    \caption{Distribution of the latent variable mean $(\mu_0, \mu_1, \mu_2)$ and the polymer parameters $(\kappa, f, \gamma L)$ in the latent space. (a) Histogram of the three latent variable of the training set. (b) Distribution of the bending modulus $\kappa$ in the latent space. (c) Distribution of the stretching force $f$ in the latent space. (d) Distribution of the steady shear $\gamma L$ in the latent space.}
    \label{fig:latent_mu_distribution}
\end{figure}

Fig.~\ref{fig:latent_mu_distribution}(a) shows the normalized distribution of the latent mean variables $\mu$ of the training set. Even without an explicit Kullback–Leibler divergence\cite{odaibo2019tutorial} term in the loss function, the latent means $\mu$ approximate a Gaussian distribution, indicating that the network naturally organizes the latent space in to a continuous and well-structured representation. As shown in Fig.~\ref{fig:latent_mu_distribution}(b)-(d), where we plot the distribution of the polymer parameters $(\kappa, f, \gamma L)$ in the $(\mu_0,\mu_1,\mu_2)$ space, the distribution of the bending modulus, stretching force and shear are well-structures and showing continuous variation in the latent space volume, which indicate the feasibility for precise mapping between the polymer parameters and the latent variables, which lead to restructure of the scattering function directly from the polymer parameters.

\begin{figure}
    \centering
    \includegraphics[width=\linewidth]{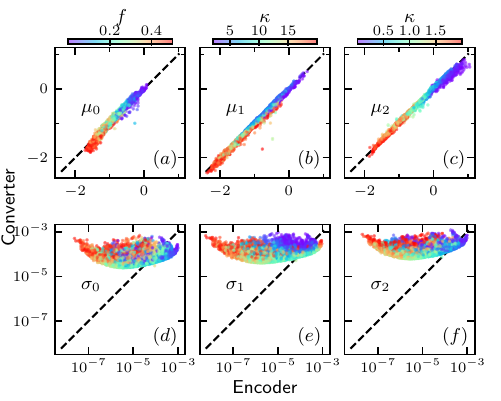}
    \caption{Comparison of the latent variable from the decoder and converter. (a)-(c) Latent variable mean $\mu_0,\mu_1,\mu_2$, colored by stretching force $f$, bending modulus $\kappa$ and steady shear $\gamma L$, respectively. (d)-(e) Latent variable standard deviation $\sigma_0, \sigma_1, \sigma_2$ with same color map as (a)-(c), respectively.}
    \label{fig:latent_comparison}
\end{figure}

Fig.~\ref{fig:latent_comparison} shows a comparison between the latent variables calculated by the encoder using the scattering function $I_{xy}(\vb{Q})$, and those derived by the Converter 1 using the polymer parameters $(\gamma, f, \gamma L)$. The means $\mu$ agrees very well as shown in Fig.~\ref{fig:latent_comparison}(a)-(c). Fig.~\ref{fig:latent_comparison}(d)-(f) shows the standard deviation $\sigma$ are misaligned, but their values are overall two order of magnitude smaller than the mean $\mu$, thus do not hinder the ability for the converting process. The $(\mu_0, \sigma_0)$, $(\mu_1,\sigma_1)$ and $(\mu_2,\sigma_2)$ in Fig.~\ref{fig:latent_comparison} are color coded by $f$, $\kappa$, and $\gamma L$, respectively. The color coding for representation are manually chosen based on the distribution of $(\gamma, f, \gamma L)$ in the $(\mu_0,\mu_1, \mu_2)$ space, as these directions appear to coincide with the chosen polymer parameters.

\subsection{Extracting polymer parameters from scattering}
In principle, the generator (comprising Converter 1 and the decoder) and the inferrer (comprising the encoder and Converter 2) enable a bidirectional mapping between the scattering function $I_{xz}(\vb{Q})$ and the polymer parameters $(\kappa, f, \gamma)$. In practice, since $I_{xz}(\vb{Q})$ can be experimentally measured using X-ray or neutron scattering techniques, extracting the model-specific polymer parameters from these measurements is highly valuable. This process not only deepens our understanding of the polymer sample’s structural characteristics but also facilitates more informed model-based material design and engineering. Both the generator and inferrer can help extracting the polymer parameters from the experimental $I^{exp}_{xz}(\vb{Q})$. The generator is a fast scattering function calculator that shortcut the MC simulation and directly output the scattering function from the polymer parameters, which can then be utilized to perform a least square fitting algorithm to find the $(\kappa, f, \gamma)$ that generate the $I'_{xz}(\vb{Q})$ that minimize $\left[\log_{10}I'_{xz}(\vb{Q})-\log_{10}'I^{exp}_{xz}(\vb{Q})\right]^2$. Meanwhile, it is straight forward to use the inferrer for parameter extraction since it can map the $I^{exp}_{xz}(\vb{Q})$ to $(\kappa, f, \gamma)$ in one forward pass.

\begin{figure}[!h]
    \centering
    \includegraphics[width=\linewidth]{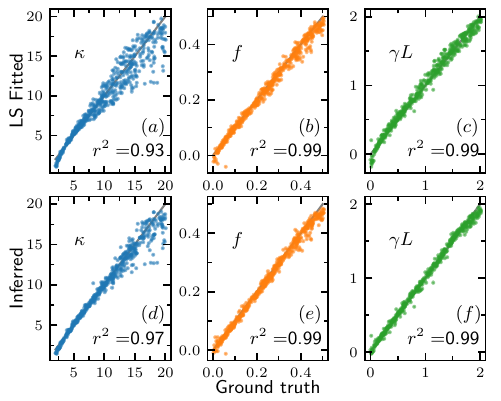}
    \caption{Comparison between the polymer parameter extracted from the scattering function and the MC input ground truth. (a) Bending modulus $\kappa$ using least-squares fitting. (b) Stretching force $f$ using least-squares fitting. (c) Steady shear $\gamma L$ using least-squares fitting. (d) Bending modulus $\kappa$ using direct inference. (e) Stretching force $f$ using direct inference. (f) Steady shear $\gamma L$ using direct inference.}
    \label{fig:fit_and_inference}
\end{figure}

Fig.~\ref{fig:fit_and_inference}(a)-(c) shows the results of least-squares fitting using the generator on the test data. The fitted polymer parameters generally agree with the MC simulation ground truth. The fitting for bending modulus $\kappa$ get slight worse at the high end with overall $r^2=0.93$, while the fitting for stretch $f$ and shear $\gamma$ achieves $r^2=0.99$ values. Meanwhile, Fig.~\ref{fig:fit_and_inference}(d)-(f) compares the polymer parameters inferred directly from the scattering function with the ground truth. Overall all of the three parameters are well-inferred, with the r-square value of the bending modulus $\kappa$ reaching $r^2=0.97$, and the stretch $f$ and shear $\gamma$ reaches $r^2=0.99$.

While the least square fit with generator and the direct inference provide equal quality of the polymer parameter extraction, the direct inference is much more efficient. Tab.~\ref{tab:run_time} compare the run time for these two method when applied on the entire test set. We test these methods on both a CPU (AMD EPYC 9334 32-core processor) and a GPU (NVIDIA RTX A6000). Overall the GPU performs the same extraction faster than the CPU. More importantly, the direct inference is 1,261 times faster on the GPU and 3,160 times faster on the CPU.

\begin{table}[!h]
    \centering
    \begin{tabular}{|c|c|c|}
    \hline
     Time  & least-square fit &  direct inference \\ \hline
      AMD EPYC 9334  & 2085.4s$^*$  &  0.66s \\ \hline 
      NVIDIA RTX A6000 &  681.43s & 0.54s \\ \hline
    \end{tabular}
    \caption{Running time for finding all of the polymer parameter from the test data using two methods on two platforms. LS- fitting target loss $5\times 10^{-4}$ and maximum iteration 2000. $^*$The AMD CPU time for the least squares fit is estimated by running on 1/10 of the full dataset and then scaled by a factor of 10 for comparison.}
    \label{tab:run_time}
\end{table}

\section{Summary}

In this work, we present a deep learning approach for analyzing two-dimensional scattering data from polymer systems using a VAE. This represents the first application of the VAE methodology to 2D scattering, enabling both accurate reconstruction of scattering patterns and direct extraction of polymer parameters. A key innovation is the introduction of an inferrer network that maps the scattering function directly to polymer parameters, achieving extraction more than 1,000 times faster than the least-squares fitting method that uses the generator. Moreover, our approach is more generalized than GPR and eliminates the need for manual hyperparameter tuning. These results highlight the potential of deep learning techniques for rapid, automated analysis of scattering data, paving the way for more efficient and scalable studies in polymer physics and materials design.

We generate the training data using our previously developed off-lattice MC simulation to avoid the orientational bias inherent in lattice models. Different polymer parameters (bending modulus $\kappa$, stretching force $f$, and steady shear $\gamma$) affect the conformation and scattering function of the polymer in distinct ways. By training the VAE, we compress the scattering function into a three-dimensional latent space and confirm the feasibility of parameter extraction by demonstrating a distinct distribution of polymer parameters in that space. Next, we train two converter networks that link the polymer parameters $(\kappa, f, \gamma)$ to the scattering function $I_{xy}(\vb{Q})$, thereby yielding both a generator and an inferrer. The generator is coupled with a least-squares fitting procedure to optimize the polymer parameters for a given $I_{xy}(\vb{Q})$ input—requiring up to 2,000 iterations—while the inferrer directly outputs the inferred parameters in a single pass. Due to its simplicity, the inferrer is approximately 1,261 times faster on our testing GPU and 3,160 times faster on the CPU than the generator approach.

The versatility and flexibility of this deep learning approach open several avenues for future research. First, while our current network employs a three-dimensional latent space to match the number of polymer parameters for ease of illustration, it is straightforward to increase the latent dimensionality to potentially achieve improved fitting and inference results. Second, our method can be extended to other scattering models and materials, such as charged polymers,\cite{ding2025charge} polymer brushes\cite{feng2018polymer} and polymer melt,\cite{kremer1990dynamics} as well as to time-dependent scattering data.\cite{smith2018dynamic} Third, it is essential to validate our approach experimentally by carrying out RheoSANS experiments to measure the scattering data from the polymer system discussed here, thereby testing the method on real experimental data. Finally, while we demonstrate the application of our neural network architecture using a SAS data, it is general enough to be applied onto spectrum analysis in other domains, for example the dynamic structure factor of active liquid interface\cite{adkins2022dynamics, zhao2024asymmetric}.

\section{Data Availability}
The code for MC simulation and ML analysis are available at \href{https://github.com/ljding94/Semiflexible_Polymer}{Semiflexible Polymer}.

\begin{acknowledgments}
This research used resources at the Spallation Neutron Source and the Center for Nanophase Materials Sciences, US Department of Energy (DOE) Office of Science User Facilities operated by Oak Ridge National Laboratory. This research was sponsored by the Laboratory Directed Research and Development Program of Oak Ridge National Laboratory, managed by UT-Battelle, LLC, for the US DOE. Computations used resources of the Oak Ridge Leadership Computing Facility, which is supported by the DOE Office of Science under contract No. DE-AC05-00OR22725. Application of machine learning to soft matter was supported by the US DOE, Office of Science, Office of Basic Energy Sciences Data, Artificial Intelligence and Machine Learning at DOE Scientific User Facilities Program under award No. 34532.
\end{acknowledgments}


\bibliography{reference}

\end{document}